\documentclass[sigconf]{acmart}

\usepackage{booktabs} 
\usepackage{lipsum}
\usepackage{multirow}
\usepackage{array}
\usepackage{courier}
\usepackage{mathtools}
\usepackage{caption}
\usepackage[font=small]{subcaption}
\usepackage{amsfonts}
\usepackage{tikz}
\usepackage{mathtools}

\DeclarePairedDelimiter\norm{\lVert}{\rVert}%
\copyrightyear{2018} 
\acmYear{2018} 
\setcopyright{acmcopyright}
\acmConference[WSDM 2018]{WSDM 2018: The Eleventh ACM International Conference on Web Search and Data Mining }{February 5--9, 2018}{Marina Del Rey, CA, USA}
\acmBooktitle{WSDM 2018: WSDM 2018: The Eleventh ACM International Conference on Web Search and Data Mining , February 5--9, 2018, Marina Del Rey, CA, USA}
\acmPrice{15.00}
\acmDOI{10.1145/3159652.3159664}
\acmISBN{978-1-4503-5581-0/18/02}

\begin{document}
\title{Hyperbolic Representation Learning for Fast and Efficient Neural Question Answering} 

\author{Yi Tay}
\affiliation{%
  \institution{Nanyang Technological University}
}
\email{ytay017@e.ntu.edu.sg}

\author{Luu Anh Tuan}
\affiliation{%
 \institution{Institute for Infocomm Research}
}
\email{at.luu@i2r.a-star.edu.sg}

\author{Siu Cheung Hui}
\affiliation{%
  \institution{Nanyang Technological University}
}
\email{asschui@ntu.edu.sg}

\begin{abstract}

The dominant neural architectures in question answer retrieval are based on recurrent or convolutional encoders configured with complex word matching layers. Given that recent architectural innovations are mostly new word interaction layers or attention-based matching mechanisms, it seems to be a well-established fact that these components are mandatory for good performance. Unfortunately, the memory and computation cost incurred by these complex mechanisms are undesirable for practical applications. As such, this paper tackles the question of whether it is possible to achieve competitive performance with simple neural architectures. We propose a simple but novel deep learning architecture for fast and efficient question-answer ranking and retrieval. More specifically, our proposed model, \textsc{HyperQA}, is a parameter efficient neural network that outperforms other parameter intensive models such as Attentive Pooling BiLSTMs and Multi-Perspective CNNs on multiple QA benchmarks. The novelty behind \textsc{HyperQA} 
is a pairwise ranking objective that models the relationship between question and answer embeddings in Hyperbolic space instead of Euclidean space. This empowers our model with a self-organizing ability and enables automatic discovery of latent hierarchies while learning embeddings of questions and answers. Our model requires no feature engineering, no similarity matrix matching, no complicated attention mechanisms nor over-parameterized layers and yet outperforms and remains competitive to many models that have these functionalities on multiple benchmarks.

 \end{abstract}

\keywords{Question Answering, Deep Learning, Learning to Rank}

\maketitle

\section{Introduction}
Neural ranking models are commonplace in many modern question answering (QA) systems \cite{DBLP:conf/sigir/SeverynM15,DBLP:conf/naacl/HeL16}. In these applications,
the problem of question answering is concerned with learning to rank candidate answers in response to questions. Intuitively, this is reminiscent of document retrieval albeit with shorter text which aggravates the long standing problem of lexical chasm \cite{DBLP:conf/sigir/BergerCCFM00}. 
For this purpose, a wide assortment of neural ranking architectures have been proposed. The key and most basic intuition pertaining to many of these models are as
follows: Firstly, representations of questions and answers are first learned via a neural encoder such as the long short-term memory (LSTM) \cite{hochreiter1997long} network or convolutional neural network (CNN). Secondly, these representations of questions and answers are composed by an interaction function to produce an overall matching score. 

The design of the interaction function between question and answer representations lives at the heart of deep learning QA research. While it is simply possible to combine QA representations with simple feed forward neural networks or other composition functions \cite{DBLP:conf/ijcai/QiuH15,DBLP:conf/sigir/TayPLH17}, a huge bulk of recent work is concerned with designing novel word interaction layers that model the relationship between the words in the QA pairs. For example, similarity matrix based matching \cite{DBLP:conf/aaai/WanLGXPC16}, soft attention alignment \cite{DBLP:conf/emnlp/ParikhT0U16} and attentive pooling \cite{DBLP:journals/corr/SantosTXZ16} are highly popular techniques for improving the performance of neural ranking models. Apparently, it seems to be well-established that grid-based matching is essential to good performance. Notably, these new innovations come with trade-offs such as huge computational cost that lead to significantly longer training times and also a larger memory footprint. Additionally, it is good to consider that the base neural encoder employed also contributes to the computational cost of these neural ranking models, e.g., LSTM networks are known to be over-parameterized and also incur a parameter and runtime cost of quadratic scale. It also seems to be a well-established fact that a neural encoder (such as the LSTM, Gated Recurrent Unit (GRU), CNN, etc.) must be first selected for learning individual representations of questions and answers and is generally treated as mandatory for good performance.

In this paper, we propose an extremely simple neural ranking model for question answering that achieves highly competitive results on several benchmarks with only a fraction of the runtime and only 40K-90K parameters (as opposed to millions). Our neural ranking models the relationships between QA pairs in Hyperbolic space instead of Euclidean space. Hyperbolic space is an embedding space with a constant negative curvature in which the distance towards the border is increasing exponentially. Intuitively, this makes it suitable for learning embeddings that reflect a natural hierarchy (e.g., networks, text, etc.) which we believe might benefit neural ranking models for QA. Notably, our work is inspired by the recently incepted Poincar\'e embeddings \cite{DBLP:journals/corr/NickelK17} which demonstrates the effectiveness of inducing a structural (hierarchical) bias in the embedding space for improved generalization. In our early empirical experiments, we discovered that a simple feed forward neural network trained in Hyperbolic space is capable of outperforming more sophisticated models on several standard benchmark datasets. We believe that this can be attributed to two reasons. Firstly, latent hierarchies are prominent in QA. Aside from the natural hierarchy of questions and answers, conceptual hierarchies also exist. Secondly, natural language is inherently hierarchical which can be traced to power law distributions such as Zipf's law \cite{ravasz2003hierarchical}. The key contributions in this paper are as follows:

\begin{itemize}

\item We propose a new neural ranking model for ranking question answer pairs. For the first time, our proposed model, \textsc{HyperQA}, performs matching of questions and answers in Hyperbolic space. To the best of our knowledge, we are the first to model QA pairs in Hyperbolic space. While hyperbolic geometry and embeddings have been explored in the domains of complex networks or graphs \cite{DBLP:journals/corr/abs-1006-5169}, our work is the first to investigate the suitability of this metric space for question answering. 
\item \textsc{HyperQA} is an extremely fast and parameter efficient model that achieves very competitive results on multiple QA benchmarks such as TrecQA, WikiQA and YahooCQA. The efficiency and speed of \textsc{HyperQA} are attributed by the fact that we do not use any sophisticated neural encoder and have no complicated word interaction layer. In fact, \textsc{HyperQA} is a mere single layered neural network with only 90K parameters. Very surprisingly, \textsc{HyperQA} actually outperforms many state-of-the-art models such as Attentive Pooling BiLSTMs \cite{DBLP:journals/corr/SantosTXZ16,DBLP:conf/aaai/ZhangLSW17} and Multi-Perspective CNNs \cite{DBLP:conf/naacl/HeL16}. We believe that this allows us to reconsider whether many of these complex word interaction layers are really necessary for good performance. 
\item We conduct extensive qualitative analysis of both the learned QA embeddings and word embeddings. We discover several interesting properties of QA embeddings in Hyperbolic space. Due to its compositional nature, we find that our model learns to self-organize not only at the QA level but also at the word-level. Our qualitative studies enable us to gain a better intuition pertaining to the good performance of our model. 

\end{itemize}

\section{Related Work}

Many prior works have established the fact that there are mainly two key ingredients to a powerful neural ranking model. First, an effective neural encoder and second, an expressive word interaction layer. The first ingredient is often treated as a given, i.e., the top performing models always use a neural encoder such as the CNN or LSTM. In fact, many top performing models adopt convolutional encoders for sentence representation \cite{DBLP:conf/emnlp/HeGL15,DBLP:conf/ijcai/QiuH15,DBLP:conf/sigir/SeverynM15,DBLP:conf/naacl/HeL16,DBLP:conf/aaai/ZhangLSW17,shen2014latent}. The usage of recurrent models is also notable \cite{DBLP:conf/aaai/MuellerT16,DBLP:conf/sigir/SeverynM15,DBLP:conf/sigir/TayPLH17,1711.07656}.

The key component in which many recent models differ at is at the interaction layer. Early works often combined QA embeddings `as it is', i.e., representations are learned first and then combined. For example, Yu et al. \cite{DBLP:journals/corr/YuHBP14} used CNN representations as feature inputs to a logistic regression model. The end-to-end CNN-based model of Severyn and Moschitti \cite{DBLP:conf/sigir/SeverynM15} combines the CNN encoded representations of question and answer using a multi-layered perceptron (MLP). Recently, a myriad of composition functions have been proposed as well, e.g., tensor layers in Qiu et al. \cite{DBLP:conf/ijcai/QiuH15} and holographic layers in Tay et al. \cite{DBLP:conf/sigir/TayPLH17}. 

It has been recently fashionable to model the relationships between question and answer using similarity matrices. Intuitively, this enables more fine-grained matching across words in question and answer sentences. The Multi-Perspective CNN (MP-CNN) \cite{DBLP:conf/emnlp/HeGL15} compared two sentences via a wide diversity of pooling functions and filter widths aiming to capture `multi-perspectives' between two sentences. The attention based neural matching (aNMM) model of Yang et al. \cite{DBLP:conf/cikm/YangAGC16} performed soft-attention alignment by first measuring the pairwise word similarity between each word in question and answer. The attentive pooling models of Santos et al. \cite{DBLP:journals/corr/SantosTXZ16} (AP-BiLSTM and AP-CNN) utilized this soft-attention alignment to learn weighted representations of question and answer that are dependent of each other. Zhang et al. \cite{DBLP:conf/aaai/ZhangLSW17} extended AP-CNN to 3D tensor-based attentive pooling (AI-CNN). A recent work, the Cross Temporal Recurrent Network (CTRN) \cite{1711.07656} proposed a pairwise gating mechanism for joint learning of QA pairs. 

Unfortunately, these models actually introduce a prohibitive computational cost to the model usually for a very marginal performance gain. Notably, it is easy to see that similarity matrix based matching incurs a computational cost of quadratic scale. Representation ability such as dimension size of word or CNN/RNN embeddings are naturally also quite restricted, i.e., increasing any of these dimensions can cause computation or memory requirements to explode. Moreover, it is not uncommon for models such as AI-CNN or AP-BiLSTM to spend more than $30$ minutes on a single epoch on QA datasets that are only medium sized. Let us not forget that these models still have to be extensively tuned which aggravates the impracticality problem posed by some of these models. 

In this paper, we seek a new paradigm for neural ranking for QA. While many recent works try to out-stack each other with new layers, we strip down our network instead. Our work is inspired by the very recent Poincar\`e embeddings \cite{DBLP:journals/corr/NickelK17} which demonstrates the superiority and efficiency of generalization in Hyperbolic space.  Moreover, this alleviates many overfitting and complexity issues that Euclidean embeddings might face especially if the data has intrinsic hierarchical structure. It is good to note that power-law distributions, such as Zipf's law, have been known to be from innate hierarchical structure \cite{ravasz2003hierarchical}. Specifically, the defining characteristic of Hyperbolic space is a much quicker expansion relative to that of Euclidean space which makes naturally equipped for modeling hierarchical structure. The concept of Hyperbolic spaces has been applied to domains such as complex network modeling \cite{DBLP:journals/corr/abs-1006-5169}, social networks \cite{verbeek2016metric} and geographic routing \cite{DBLP:conf/infocom/Kleinberg07}. 

\begin{figure}
\centering
\begin{subfigure}{0.25\textwidth}
  \centering
  \includegraphics[width=0.4\linewidth]{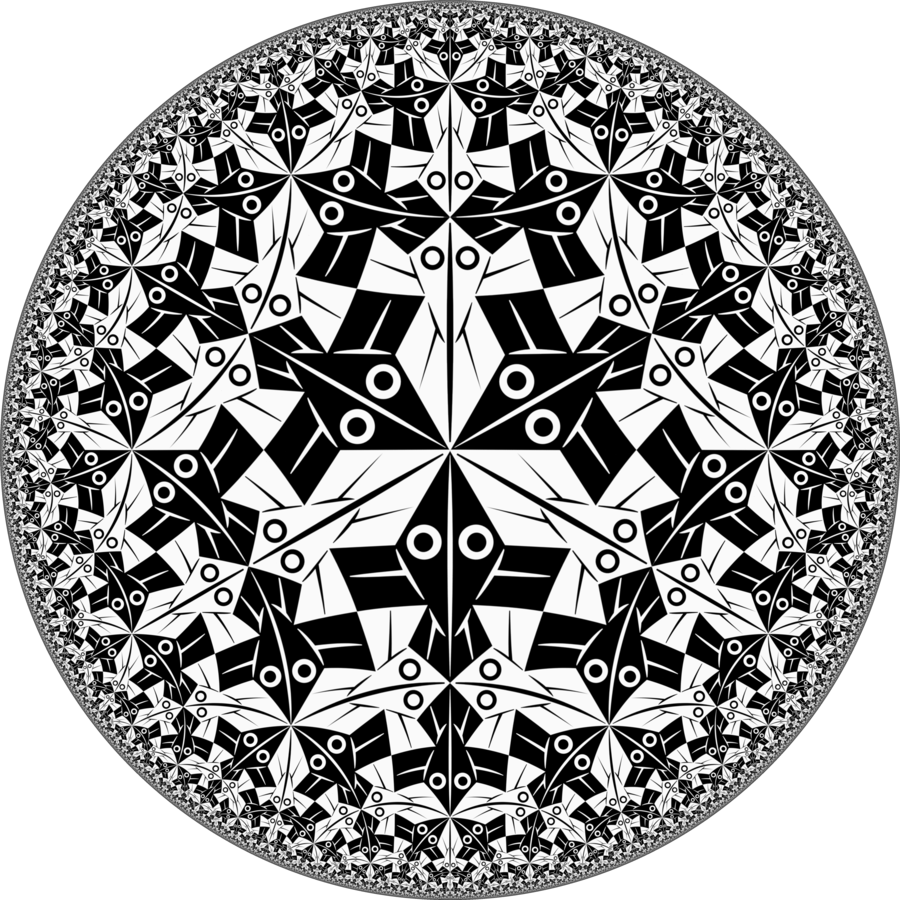}
  \caption{`Circle Limit 1' by M.C Escher}
  \label{fig:sub1}
\end{subfigure}%
\begin{subfigure}{0.25\textwidth}
  \centering
  \includegraphics[width=0.4\linewidth]{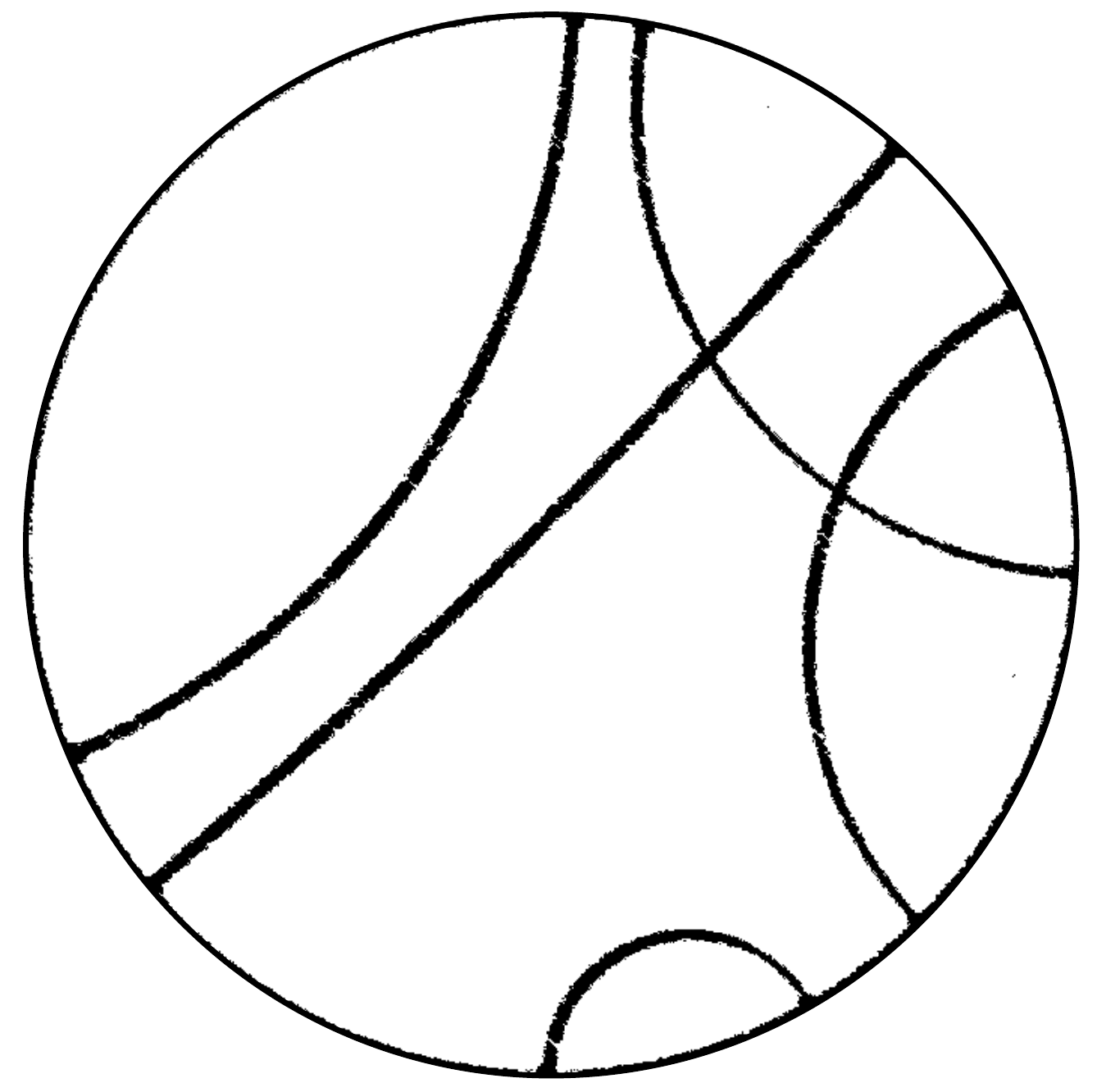}
  \caption{Hyperbolic Parallel Lines}
  \label{fig:sub2}
\end{subfigure}
\caption{Visualizations of Hyperbolic space.}
\label{fig:hyp}
\end{figure}

There are several key geometric intuitions regarding Hyperbolic spaces. Firstly, the concept of distance and area is \textit{warped} in Hyperbolic spaces. Specifically, each tile in Figure \ref{fig:hyp}(a) is of equal area in Hyperbolic space but diminishes towards zero in Euclidean space towards the boundary. Secondly, Hyperbolic spaces are \textit{conformal}, i.e., angles in Hyperbolic spaces and Euclidean spaces are identical. In Figure \ref{fig:hyp}(b), the arcs on the curve are parallel lines that are orthogonal to the boundary. Finally, hyperbolic spaces can be regarded as \textit{larger} spaces relative to Euclidean spaces due to the fact that the concept of relative distance can be expressed much better, i.e., not only does the distance between two vectors encode information but also \textit{where} a vector is placed in Hyperbolic space. This enables efficient representation learning.

In Nickel et al. \cite{DBLP:journals/corr/NickelK17}, the authors applied the hyperbolic distance (specifically, the Poincar\`e distance) to model taxonomic entities and graph nodes. Notably, our work, to the best of our knowledge, is the only work that learns QA embeddings in Hyperbolic space. Moreover, questions and answers introduce an interesting layer of complexity to the problem since QA embeddings are in fact compositions of their constituent word embeddings. On the other hand, nodes in a graph and taxonomic entities in \cite{DBLP:journals/corr/NickelK17} are already at its most abstract form, i.e., symbolic objects. As such, we believe it would be interesting to investigate the impacts of QA in Hyperbolic space in lieu of the added compositional nature.

\section{Our Proposed Approach}
\begin{figure}[ht]
\begin{center}
\includegraphics[width=0.4\textwidth]{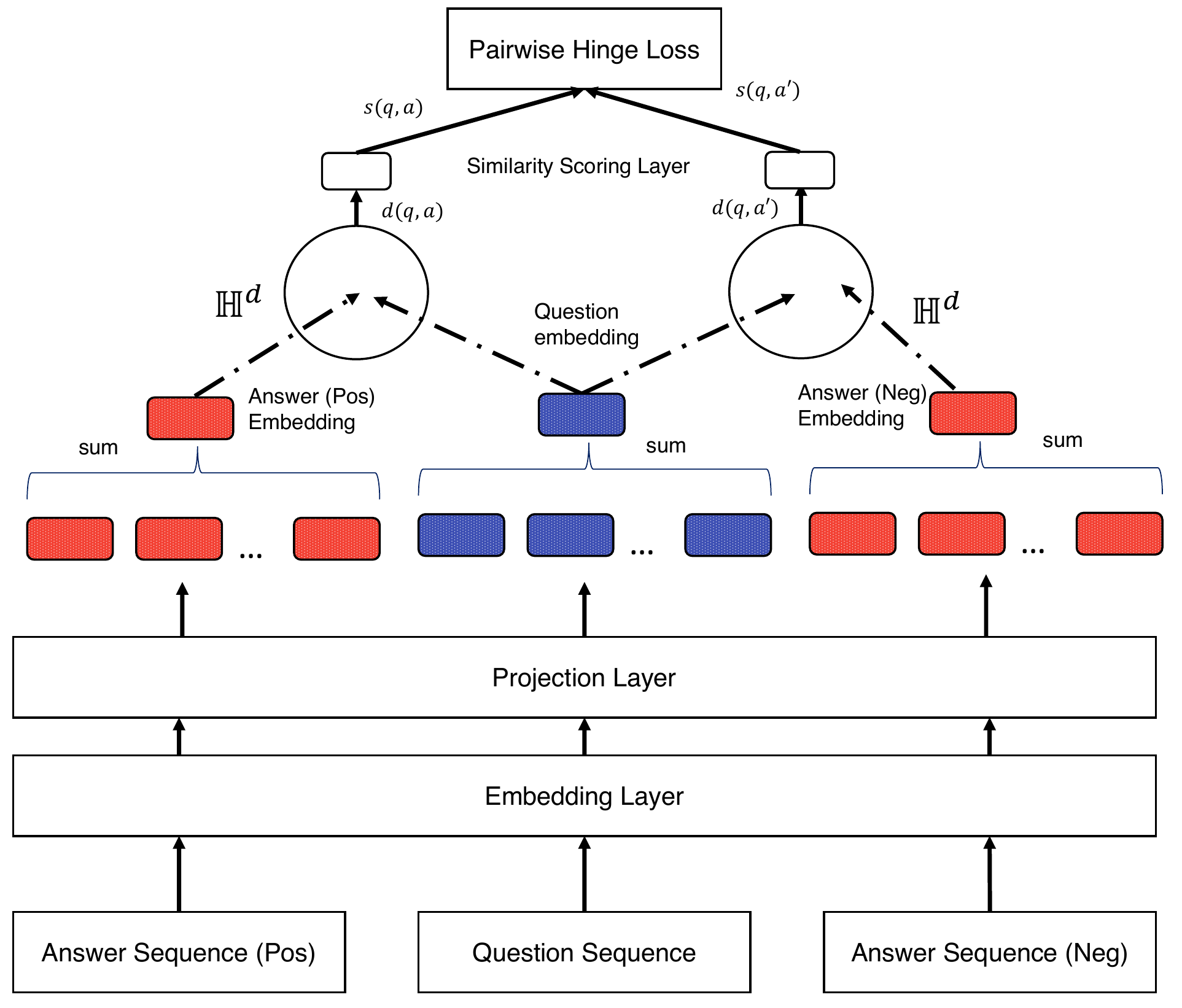}
\caption{Our proposed model architecture.}
\label{overall}
\end{center}
\end{figure}

This section outlines the overall architecture of our proposed model. Similar to many neural ranking models for QA, our network has `two' sides with shared parameters, i.e., one for question and another for answer. However, since we optimize for a pairwise ranking loss, the model takes in a positive (correct) answer and a negative (wrong) answer and aims to maximize the margin between the scores of the correct QA pair and the negative QA pair. Figure \ref{overall} depicts the overall model architecture. 
\vspace{-1em}

\subsection{Embedding Layer}
Our model accepts three sequences as an input, i.e., the question (denoted as $q$), the correct answer (denoted as $a$) and a randomly sampled corrupted answer (denoted as $a'$). Each sequence consists of $M$ words where $M_{q}$ and $M_{a}$ are predefined maximum sequence lengths for questions and answers respectively. Each word is represented as a one-hot vector (representing a word in the vocabulary). As such, this layer is a look-up layer that converts each word into a low-dimensional vector by indexing onto the word embedding matrix. In our implementation, we initialize this layer with pretrained word embeddings \cite{DBLP:conf/emnlp/PenningtonSM14}. Note that this layer is not updated during training. Instead, we utilize a projection layer that learns a task-specific projection of the embeddings. 

\subsection{Projection Layer}
In order to learn a task-specific representation for each word, we utilize a projection layer. The projection layer is essentially a single layered neural network that is applied to \textbf{each word} in all three sequences. 

\begin{equation}
x = \sigma(\textbf{W}_p \:z + b_p)
\end{equation}
where $\textbf{W}_p \in \mathbb{R}^{d \times n}$, $z \in \mathbb{R}^{n}$, $x \in \mathbb{R}^{d}$ and $\sigma$ is a non-linear function such as the rectified linear unit (ReLU). The output of this layer is a sequence of $d$ dimensional embeddings for each sequence (question, positive answer and negative answer). Note that the parameters of this projection layer are shared for both question and answer. 

\subsection{Learning QA Representations}
In order to learn question and answer representations, we simply take the sum of all word embeddings in the sequence. 
\begin{equation}
y^{*} = \sum_{i=1}^{M_*} x_{i}^{*}
\end{equation}
\label{qa_sec}
where $*=\{q,a,a'\}$. $M$ is the predefined max sequence length (specific to question and answer) and $x_1, x_2 \dots x_{M}$ are $d$-dimensional embeddings of the sequence. This is essentially the neural bag-of-words (NBoW) representation. Unlike popular neural encoders such as LSTM or CNN, the NBOW representation does not add any parameters and is much more efficient. Additionally, we constrain the question and answer embeddings to the unit ball before passing to the next layer, i.e.,  $\norm{y^*} \leq 1$. This is easily done via $y^* = \frac{y^*}{\norm{y^*}}$  when $\norm{y^*} > 1$. Note that this projection of QA embeddings onto the unit ball is mandatory and absolutely crucial for \textsc{HyperQA} to even work. 

\subsection{Hyperbolic Representations of QA Pairs}
Neural ranking models are mainly characterized by the interaction function between question and answer representations. In our work, we mainly adopt the hyperbolic\footnote{While there exist multiple models of Hyperbolic geometry such as the Beltrami-Klein model or the Hyperboloid model, we adopt the Poincar\`e ball / disk due to its ease of differentiability and freedom from constraints \cite{DBLP:journals/corr/NickelK17}.} distance function to model the relationships between questions and answers. Formally, let $\mathcal{B}^{d} = \{x \in \mathbb{R}^{d} \:|\: \norm{x} < 1 \}$ be the \textit{open} $d$-dimensional unit ball, our model corresponds to the Riemannian manifold ($\mathcal{B}^{d}, g_{x}$) and is equipped with the Riemannian metric tensor given as follows:

\begin{equation}
g_x = (\frac{2}{1-\norm{x}^2})^2 g^{E}
\end{equation}
where $g^{E}$ is the Euclidean metric tensor. The hyperbolic distance function between question and answer is defined as:
\begin{equation}
d(q,a) = arcosh (1 + 2 \frac{\norm{q-a}^2}{(1-\norm{q}^2)(1-\norm{a}^2)})
\end{equation}
where $\norm{.}$ denotes the Euclidean norm and $q,a \in \mathbb{R}^{d}$ are the question and answer embeddings respectively. Note that $arcosh$ is the inverse hyperbolic cosine function, i.e., $arcosh \: x = \ln (x + \sqrt{(x^2 -1)})$. Notably, $d(q,a)$ changes smoothly with respect to the position of $q$ and $a$ which enables the automatic discovery of latent hierarchies. As mentioned earlier, the distance increases exponentially as the norm of the vectors approaches 1. As such, the latent hierarchies of QA embeddings are captured through the norm of the vectors. From a geometric perspective, the origin can be seen as the root of a tree that branches out towards the boundaries of the hyperbolic ball. This self-organizing ability of the hyperbolic distance is visually and qualitatively analyzed in later sections. 

\subsubsection{Gradient Derivation}
Amongst the other models of Hyperbolic geometry, the hyperbolic Poincar\`e distance is differentiable. Let  The partial derivate w.r.t to $\theta$ is defined as:
\begin{equation}
\frac{\partial d(\theta, x)}{\partial \theta} 
= \frac{4}{\beta \sqrt{\gamma^2 - 1}} (\frac{\norm{x}^{2} - 2 \langle \theta, x \rangle + 1}{\alpha^2} \theta - \frac{x}{\alpha})
\end{equation}
where $\alpha = 1-\norm{\theta}^2$, $\beta = 1-\norm{x}^2$ and $\gamma = 1 + \frac{2}{\alpha \beta} \norm{\theta - x}^2$. 

\subsection{Similarity Scoring Layer}
Finally, we pass the hyperbolic distance through a linear transformation described as follows:

\begin{equation}
s(q,a) = w_f \: d(q,a) + b_f
\end{equation}
where $w_f \in \mathbb{R}^{1}$ and $b_f \in \mathbb{R}^{1}$ are scalar parameters of this layer. The performance of this layer is empirically motivated by its performance and was selected amongst other variants such as $exp(-d(q,a))$, non-linear activations such as sigmoid function or the raw hyperbolic distance.

\subsection{Optimization and Learning}
This section describes the optimization and learning process of \textsc{HyperQA}. Our model learns via a pairwise ranking loss, which is well suited for metric-based learning algorithms.  
\subsubsection{Pairwise Hinge Loss}
Our network minimizes the pairwise hinge loss which is defined as follows:
\begin{equation}
L =  \sum_{(q,a) \in \Delta_q} \sum_{(q,a') \not\in \Delta_q} max(0,s(q,a) + \lambda - s(q,a'))
\end{equation}
where $\Delta_q$ is the set of all QA pairs for question $q$, $s(q,a)$ is the score between $q$ and $a$, and $\lambda$ is the margin which controls the extent of discrimination between positive QA pairs and corrupted QA pairs. The adoption of the pairwise hinge loss is motivated by the good empirical results demonstrated in Rao et al. \cite{DBLP:conf/cikm/RaoHL16}. Additionally, we also adopt the \textit{mix sampling} strategy for sampling negative samples as described in their work. 

\subsubsection{Gradient Conversion}
Since our network learns in Hyperbolic space, parameters have to be learned via stochastic Riemannian optimization methods such as RSGD \cite{DBLP:journals/tac/Bonnabel13}. 
\begin{equation}
\theta_{t+1} = \Re_{\theta_{t}} (- \eta \nabla_{R} \: \ell (\theta_t))
\end{equation}
where $\Re_{\theta_t}$ denotes a retraction onto $\mathcal{B}$ at $\theta$. $\eta$ is the learning rate and $\nabla_{R} \: \ell (\theta_t)$ is the Riemannian gradient with respect to $\theta_t$. Fortunately, the Riemannian gradient can be easily derived from the Euclidean gradient in this case \cite{DBLP:journals/tac/Bonnabel13}. In order to do so, we can simply scale the Euclidean gradient by the inverse of the metric tensor $g_{\theta}^{-1}$. Overall, the final gradients used to update the parameters are:
\begin{equation}
\label{gradient_scale}
\nabla_{R} = \frac{(1-\norm{\theta_{t}}^2)^{2}}{4} \nabla_{E} 
\end{equation} 
Due to the lack of space, we refer interested readers to \cite{DBLP:journals/corr/NickelK17,DBLP:journals/tac/Bonnabel13} for more details. For practical purposes, we simply utilize the automatic gradient feature of TensorFlow but convert the gradients with Equation (\ref{gradient_scale}) before updating the parameters.

\section{Experiments}
This section describes our empirical evaluation and its results. 
\subsection{Datasets}
In the spirit of experimental rigor, we conduct our empirical evaluation based on four popular and well-studied benchmark datasets for question answering. 

\begin{itemize}
	\item \textbf{YahooCQA} -  This is a benchmark dataset for community-based question answering that was collected from Yahoo Answers. In this dataset, the answer lengths are relatively longer than TrecQA and WikiQA. Therefore, we filtered answers that have more than $50$ words and less than $5$ characters. The train-dev-test splits for this dataset are provided by \cite{DBLP:conf/sigir/TayPLH17}.

	\item \textbf{WikiQA} - This is a recently popular benchmark dataset \cite{DBLP:conf/emnlp/YangYM15} for open-domain question answering based on factual questions from Wikipedia and Bing search logs.
	\item \textbf{SemEvalCQA} - This is a well-studied benchmark dataset from SemEval-2016 Task 3 Subtask A (CQA). This is a real world dataset obtained from Qatar Living Forums. In this dataset, there are ten answers in each question `thread' which are marked as `Good`, `Potentially Useful' or `'Bad'. We treat `Good' as positive and anything else as negative labels. 
\item \textbf{TrecQA} - This is the benchmark dataset provided by Wang et al. \cite{DBLP:conf/emnlp/WangSM07}. This dataset was collected from TREC QA tracks 8-13 and is comprised of factoid based questions which mainly answer the `who', `what', `where', `when' and `why' types of questions. There are two versions, namely clean and raw, as noted by \cite{DBLP:conf/cikm/RaoHL16} which we evaluate our models on. 

\end{itemize}
Statistics pertaining to each dataset is given in Table \ref{tab:dataset}.

\begin{table}[H]
  \centering
  \small
    \begin{tabular}{lcccc}
    \hline
         			& YahooCQA & WikiQA & SemEvalCQA   & TrecQA\\
          \hline
    Train Qns &   50.1K  & 94        &  4.8K  & 1229 \\
    Dev Qns & 6.2K & 65      &224    &  82 \\
    Test Qns & 6.2K & 68    &  327   & 100 \\
    \hline
    Train Pairs & 253K& 5.9K   &36K   & 53 \\
    Dev Pairs &31.7K &1.1K   &2.4K  &   1.1K\\
    Test Pairs & 31.7K &  1.4K & 3.2K    & 1.5K \\
    \hline
    \end{tabular}%
    \caption{Statistics of datasets.}
  \label{tab:dataset}%
\end{table}%

\subsection{Compared Baselines}

In this section, we introduce the baselines for comparison. 

\begin{itemize}

\item \textbf{YahooCQA -}
The key competitors of this dataset are the Neural Tensor LSTM (NTN-LSTM) and HD-LSTM from Tay et al. \cite{DBLP:conf/sigir/TayPLH17} along with their implementation of the Convolutional Neural Tensor Network \cite{DBLP:conf/ijcai/QiuH15}, vanilla CNN model, and the Okapi BM-25 \cite{DBLP:conf/trec/RobertsonWJHG94} benchmark. Additionally, we also report our own implementations of QA-BiLSTM, QA-CNN, AP-BiLSTM and AP-CNN on this dataset based on our experimental setup.

\item \textbf{WikiQA -}
The key competitors of this dataset are the Paragraph Vector (PV) and PV + Cnt models \cite{le2014distributed} of Le and Mikolv, CNN + Cnt model from Yu et al. \cite{DBLP:journals/corr/YuHBP14} and LCLR (Yih et al.) \cite{DBLP:conf/acl/YihCMP13}. These three baselines are reported in the original WikiQA paper \cite{DBLP:conf/emnlp/YangYM15} which also include variations that include handcrafted features. Additional strong baselines include QA-BiLSTM, QA-CNN from \cite{DBLP:journals/corr/SantosTXZ16} along with AP-BiLSTM and AP-CNN which are attentive pooling improvements of the former. Finally, we also report the Pairwise Ranking MP-CNN from Rao et al. \cite{DBLP:conf/cikm/RaoHL16}.

\item \textbf{SemEvalCQA -}
The key competitors of this dataset are the CNN-based ARC-I/II architecture by Hu et al. \cite{DBLP:conf/nips/HuLLC14}, the Attentive Pooling CNN \cite{DBLP:journals/corr/SantosTXZ16}, Kelp \cite{DBLP:conf/semeval/FiliceCMB16} a feature engineering based SVM method, ConvKN \cite{DBLP:conf/semeval/Barron-CedenoMJ16} a combination of convolutional tree kernels with CNN and finally  AI-CNN (Attentive Interactive CNN) \cite{DBLP:conf/aaai/ZhangLSW17}, a tensor-based attentive pooling neural model. A comparison with AI-CNN (with features) is also included. 

\item \textbf{TrecQA -} 
The key competitors on the dataset are mainly the CNN model of Severyn and Moschitti (S\&M) \cite{DBLP:conf/sigir/SeverynM15}, the Attention-based Neural Matching Model (aNMM) of Yang et al. \cite{DBLP:conf/cikm/YangAGC16}, HD-LSTM (Tay et al.) \cite{DBLP:conf/sigir/TayPLH17} and Multi-Perspective CNN (MP-CNN) \cite{DBLP:conf/emnlp/HeGL15} proposed by He et al. Lastly, we also compare with the pairwise ranking adaption of MP-CNN (Rao et al.) \cite{DBLP:conf/cikm/RaoHL16}. Additionally and due to long standing nature of this dataset, there have been a huge number of works based on traditional feature engineering approaches \cite{DBLP:conf/emnlp/WangSM07,DBLP:conf/naacl/HeilmanS10a,DBLP:conf/sigir/SeverynMTBR14,DBLP:conf/naacl/YaoDCC13} which we also report. For the clean version of this dataset, we also compare with AP-CNN and QA-BiLSTM/CNN \cite{DBLP:journals/corr/SantosTXZ16}.

\end{itemize}
Since the training splits are standard, we are able to directly report the results from the original papers.

\subsection{Evaluation Protocol}
This section describes the key evaluation protocol / metrics and implementation details of our experiments. 
\subsubsection{Metrics}
We adopt a dataset specific evaluation protocol in which we follow the prior work in their evaluation protocols. Specifically, TrecQA and WikiQA adopt the Mean Reciprocal Rank (MRR) and MAP (Mean Average Precision) metrics which are commonplace in IR research. On the other hand, YahooCQA and SemEvalCQA evaluate on MAP and Precision@1 (abbreviated P@1) which is determined based on whether the top predicted answer is the ground truth. For all competitor methods, we report the performance results from the original paper. 

\subsubsection{Training Time \& Parameter Size}
Additionally, we report the parameter size and runtime (seconds per epoch) of selected models.  We selectively re-implement some of the key competitors with the best performance and benchmark their training time on our machine/GPU (a single Nvidia GTX1070). For reporting the parameter size and training time, we try our best to follow the hyperparameters stated in the original papers. As such, the same model can have different training time and parameter size on different datasets. 

\subsubsection{Hyperparameters}
\textsc{HyperQA} is implemented in TensorFlow. We adopt the AdaGrad \cite{DBLP:journals/jmlr/DuchiHS11} optimizer with initial learning rate tuned amongst $\{0.2,0.1,0.05,0.01\}$. The batch size is tuned amongst $\{50,100,200\}$. Models are trained for $25$ epochs and the model parameters are saved each time the performance on the validation set is topped. The dimension of the projection layer is tuned amongst $\{100,200,300,400\}$. L2 regularization is tuned amongst $\{0.001, 0.0001, 0.00001\}$. The negative sampling rate is tuned from $2$ to $8$. Finally, the margin $\lambda$ is tuned amongst $\{1,2, 5,10,20\}$. For TrecQA, WikiQA and YahooCQA, we initialize the embedding layer with GloVe \cite{DBLP:conf/emnlp/PenningtonSM14} and use the version with $d=300$ and trained on 840 billion words. For SemEvalCQA, we train our own Skipgram model using the unannotated corpus provided by the task.  In this case, the embedding dimension is tuned amongst $\{100,200,300\}$. Embeddings are not updated during training. For the SemEvalCQA dataset, we concatenated the raw QA embeddings before passing into the final layer since we found that it improves performance.

\subsection{Results and Analysis}
In this section, we present our empirical results on all datasets. For all reported results, the best result is in boldface and the second best is underlined.

\subsubsection{Experimental Results on WikiQA}
Table \ref{tab:wiki_results} reports our results on the WikiQA dataset. Firstly, we observe that \textsc{HyperQA} outperforms a myriad
of complex neural architectures. Notably, we obtain a clear performance gain of $2\%-3\%$ in terms of MAP/MRR against models such as AP-CNN or
AP-BiLSTM. Our model also outperforms MP-CNN which is severely equipped with parameterized word matching mechanisms.  We achieve competitive results
relative to the Rank MP-CNN. Finally, \textsc{HyperQA} is extremely efficient and fast, clocking 2s per epoch compared to 33s per epoch for Rank MP-CNN. The parameter cost
is also 90K vs 10 million which is a significant improvement. 


\begin{table}[ht]
  \centering
\small
    \begin{tabular}{ccccc}
    \hline
       Model   & MAP   & MRR   & \#Params & Time \\
          \hline
    PV  & 0.511 & 0.516 & -     & - \\
    PV + Cnt  & 0.599 & 0.609 & -     & - \\
    LCLR & 0.599 & 0.609 & -     & - \\

    CNN + Cnt & 0.652 & 0.665 & -     & - \\
    QA-BiLSTM  (Santos et al.)& 0.656 & 0.670 & -    & - \\
    QA-CNN  (Santos et al.)& 0.670 & 0.682 & -    & - \\
    AP-BiLSTM  (Santos et al.)& 0.671 & 0.684 & -     & - \\
    AP-CNN (Santos et al.) & 0.688 & 0.696 & -    & - \\
    MP-CNN (He et al.) & 0.693 & 0.709 & 10.0M   & 35s\\
    Rank MP-CNN (Rao et al.) & \underline{0.701} & \underline{0.718} & 10.0M  & 33s \\

   \hline
    \textsc{HyperQA} (This work) & \textbf{0.712} & \textbf{0.727} & 90K   & 2s \\
    \hline
    \end{tabular}%

      \caption{Experimental results on WikiQA.}
  \label{tab:wiki_results}%
\end{table}%
\subsubsection{Experimental Results on YahooCQA}

Table \ref{tab:yahooQAresults} reports the experimental results on YahooCQA. First, we observe that \textsc{HyperQA} outperforms AP-BiLSTM and AP-CNN significantly. Specifically, we outperform AP-BiLSTM, the runner-up model by $6\%$ in terms of MRR and $10\%$ in terms of MAP. Notably, \textsc{HyperQA} is 32 times faster than AP-BiLSTM and has $20$ times less parameters. Our approach shows that complicated attentive pooling mechanisms are not necessary for good performance. 

\begin{table}[ht]
  \centering
\small
    \begin{tabular}{lcccc}
    \hline
          
         Model & P@1   & MRR   & \# Params & Time \\
          \hline
    Random Guess & 0.200 & 0.457 & -      & - \\
    BM-25 & 0.225 & 0.493 &  -     & - \\

    CNN  & 0.413 & 0.632 &  -     & - \\
    CNTN (Qiu et al.)  & 0.465 & 0.632 & -      & - \\
    LSTM & 0.465 & 0.669 &  -     & - \\
    NTN-LSTM (Tay et al.)  & 0.545 & 0.731 & -     & - \\
    HD-LSTM (Tay et al.)  & 0.557 & 0.735 & -      & - \\

    QA-BiLSTM (Santos et al.)& 0.508 &0.683 &1.40M & 440s \\
    QA-CNN (Santos et al.) & 0.564& 0.727&  90.9K & 60s \\
    AP-CNN (Santos et al.) & 0.560& 0.726& 540K & 110s \\
    AP-BiLSTM (Santos et al.) & \underline{0.568}& \underline{0.731}& 1.80M& 640s \\
    \hline
    \textsc{HyperQA} (This work) & \textbf{0.683} & \textbf{0.801} & 90.0K   & 20s \\
    \hline
    \end{tabular}%
     \caption{Experimental results on YahooCQA.}
  \label{tab:yahooQAresults}%
\end{table}

\subsubsection{Experimental Results on SemEvalCQA}
Table \ref{tab:results_se} reports the experimental results on SemEvalCQA. Our proposed approach achieves highly competitive performance on this dataset. Specifically, we have obtained the best P@1 performance overall, outperforming
the state-of-the-art AI-CNN  model by $3\%$ in terms of P@1. The performance of our model on MAP is marginally short from the best performing model. Notably, AI-CNN has benefited from external handcrafted features. As such, comparing AI-CNN (w/o features) with \textsc{HyperQA} shows that our proposed model is a superior neural ranking model. Next, we draw the readers attention to the time cost of AI-CNN. The training time per epoch is $\approx 3250s$ per epoch which is about $300$ times longer than our model. AI-CNN is extremely cost prohibitive, i.e., attentive pooling is already very expensive and yet AI-CNN performs 3D attentive pooling. Evidently, its performance can be easily superseded in a much smaller training time and parameter cost. This raises questions about the effectiveness of the 3D attentive pooling mechanism.

\begin{table}[ht]
\small
  \centering

    \begin{tabular}{lcccc}
    \hline
    Model & P@1   & MAP   & \#Params & Time \\
    \hline
    ARC-I (Hu et al.) & 0.741 & 0.771 &  -     & - \\
    ARC-II (Hu et al.) & 0.753 & 0.780 &  -     & - \\
    AP-CNN (Santos et al.)   & 0.755 & 0.771 &   -    & - \\
    Kelp  (Filice et al.) & 0.751 & 0.792 &   -    & -  \\
    ConvKN (Barr{\'{o}}n{-}Cede{\~{n}}o et al.)& 0.755 & 0.777 &  -     & -  \\
    AI-CNN (Zhang et al.)  & 0.763 & 0.792 &  140K     &  3250s\\
    AI-CNN + Feats (Zhang et al.) & \underline{0.769} & \textbf{0.801} &  140K      & 3250s \\
    \hline
    HyperQA (This work) & \textbf{0.809} & \underline{0.795} & 45K    &  10s \\
    \hline
    \end{tabular}%

    \caption{Experimental results on SemEvalCQA.}
  \label{tab:results_se}%
\end{table}%

\begin{table}[ht]
  \centering
  \small
    \begin{tabular}{lcccc}
    \hline
    Model & MAP   & MRR   & \# Params & Time \\
    \hline
    Wang et al. (2007) & 0.603 & 0.685 & -     & - \\
    Heilman et al. (2010) & 0.609 & 0.692 & -     & - \\
    Wang et al. (2010)  & 0.595 & 0.695 & -     & - \\
    Yao (2013) & 0.631 & 0.748 & -     & - \\
    Severyn and Moschitti (2013) & 0.678 & 0.736 & -     & - \\
    Yih et al (2014) & 0.709 & 0.770  & -     & - \\
    \hline
    CNN (Yu et al) & 0.711 & 0.785 & -     & - \\
    BLSTM + BM25 (Wang \& Nyberg) & 0.713 & 0.791 & -     & - \\
    CNN (Severyn \&  Moschitti)   & 0.746 & 0.808 & -     & - \\
    aNMM (Yang et al.)  & 0.750 & 0.811 & -     & - \\
    HD-LSTM (Tay et al.) & 0.750 & 0.815 & -     & - \\
    MP-CNN  (He et al.) & 0.762 & 0.822 & 10.0M   & 141s \\
    Rank MP-CNN (Rao et al.) & \textbf{0.780}  & \textbf{0.830} & 10.0M  &  130s\\
    \hline
    \textsc{HyperQA} (This work) & \underline{0.770} & \underline{0.825} & \textbf{90K}   & 12s \\
    \hline
    \end{tabular}%
    \caption{Experimental results on TrecQA (raw). Feature engineering and deep learning approaches are separated by the middle line.}
  \label{tab:trec_results}%
\end{table}%

\begin{table}[htbp]
  \centering
  \small
    \begin{tabular}{lcccc}
    \hline
    Model & MAP   & MRR   &  \# Params     &  Time\\
    \hline
    QA-LSTM /  CNN (Santos et al.) & 0.728 & 0.832 &  -     & - \\
    AP-CNN (Santos et al.) & 0.753 & 0.851 &    -   & - \\
    MP-CNN (He et al.) & 0.777 & 0.836 & 10M   & 141 \\
    Rank MP-CNN (Rao et al.) & \textbf{0.801} & \textbf{0.877} & 10M   & 130s \\
    \hline
    HyperQA & \underline{0.784} & \underline{0.865} & 90K      & 12s \\
    \hline
    \end{tabular}%
     \caption{Experimental results on TrecQA (clean)}
  \label{tab:trec_results_2}%
\end{table}%

\subsubsection{Experimental Results on TrecQA}
Table \ref{tab:trec_results} reports the results on TrecQA (raw). \textsc{HyperQA} achieves very competitive performance on both MAP and MRR metrics. Specifically, \textsc{HyperQA} outperforms the basic CNN model of (S\&M) by  $2\%-3\%$ in terms of MAP/MRR. Moreover, the CNN (S\&M) model uses handcrafted features which \textsc{HyperQA} does not require. Similarly, the aNMM model and HD-LSTM also benefit from additional features but are outperformed by \textsc{HyperQA}. \textsc{HyperQA} also outperforms MP-CNN but is around $10$ times faster and has $100$ times less parameters. MP-CNN consists of a huge number of filter banks and utilizes heavy parameterization to match multiple perspectives of questions and answers. On the other hand, our proposed \textsc{HyperQA} is merely a single layered neural network with 90K parameters and yet outperforms MP-CNN. Similarly, Table \ref{tab:trec_results_2} reports the results on TrecQA (clean). Similarly, \textsc{HyperQA} also outperforms MP-CNN, AP-CNN and QA-CNN. On both datasets, the performance of \textsc{HyperQA} is competitive to Rank MP-CNN.

\subsubsection{Overall analysis}
Overall, we summarize the key findings of our experiments.
\begin{itemize}
\item It is possible to achieve very competitive performance with small parameterization, and no word matching or interaction layers. \textsc{HyperQA} outperforms complex models such as MP-CNN and AP-BiLSTM on multiple datasets. 
\item The relative performance of \textsc{HyperQA} is significantly better on large datasets, e.g., YahooCQA (253K training pairs) as opposed to smaller ones like WikiQA (5.9K training pairs). We believe that this is due to the fact that Hyperbolic space is \textit{seemingly} larger than Euclidean space. 
\item \textsc{HyperQA} is extremely fast and trains at $10-20$ times faster than complex models like MP-CNN. Note that if CPUs are used instead of GPUs (which speed convolutions up significantly), this disparity would be significantly larger. 
\item Our proposed approach does not require handcrafted features and yet outperforms models that benefit from them. This is evident on all datasets, i.e., \textsc{HyperQA} outperforms CNN model with features (TrecQA and WikiQA) and AI-CNN + features on SemEvalCQA. 
\end{itemize}

\begin{table}[htbp]
  \centering
\small
    \begin{tabular}{lccc}
    \hline
    Ours against & Performance & Params & Speed \\
    \hline
    AP-BiLSTM & 1-7\% better & 20x less & 32 x faster \\
    AP-CNN & 1-12\% better & Same  & 3x faster \\
    AI-CNN & Competitive & 3x less & 300x faster \\
    MP-CNN & 1-2\% better & 100x less & 10x faster \\
    Rank MP-CNN & Competitive & 100x less & 10x faster \\
    \hline
    \end{tabular}%
    \caption{Overall comparison of \textsc{HyperQA} against other state-of-the-art models.}
  \label{tab:addlabel}%
\end{table}%

\subsection{Effects of QA Embedding Size}
\begin{figure}[H] 
\begin{center}
\includegraphics[width=0.35\textwidth]{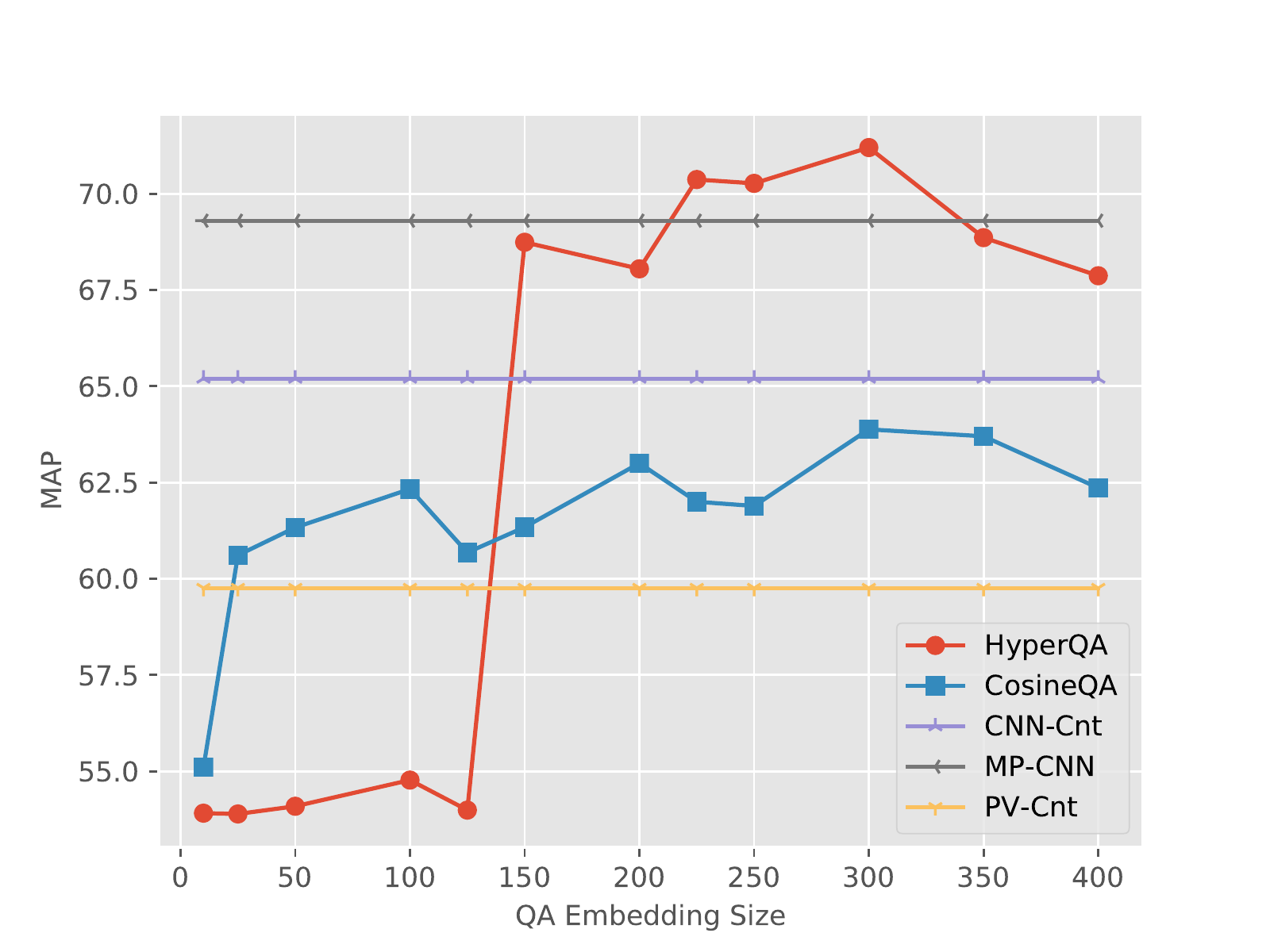}
\caption{Effects of QA embedding size on WikiQA.}
\label{emb_size}
\end{center}
\end{figure}
\vspace{-1em}

In this section, we study the effects of the QA embedding size on performance. Figure \ref{emb_size} describes the relationship between QA embedding size ($d$) and MAP on the WikiQA dataset. Additionally, we include a simple baseline (CosineQA) which is exactly the same as \textsc{HyperQA} but uses cosine similarity instead of hyperbolic distance. The MAP scores of three other reported models (MP-CNN, CNN-Cnt and PV-Cnt) are also reported for reference. Firstly, we notice the disparity between \textsc{HyperQA} and CosineQA in terms of performance. This is also observed across other datasets but is not reported due to the lack of space. While CosineQA maintains a stable performance throughout embedding size, the performance of \textsc{HyperQA} rapidly improves at $d>150$. In fact, the performance of \textsc{HyperQA} at $d=150$ (45K parameters) is already similar to the Multi-Perspective CNN \cite{DBLP:conf/emnlp/HeGL15} which contains 10 million parameters. Moreover, the performance of \textsc{HyperQA} outperforms MP-CNN with $d=250$-$300$.

\section{Discussion and Analysis}
This section delves into qualitative analysis of our model and aims to investigate the following research questions:
\begin{enumerate}
\item \textbf{RQ1:} Is there any hierarchical structure learned in the QA embeddings? How are QA embeddings organized in the final embedding space of \textsc{HyperQA}?
\item \textbf{RQ2:} What are the impacts of embedding compositional embeddings in hyperbolic space? Is there an impact on the constituent word embeddings?
\item \textbf{RQ3:} Are we able to derive any insight about how word interaction and matching happens in \textsc{HyperQA}?
\end{enumerate}

\begin{table*}[htbp]
  \centering
  \small
  
    \begin{tabular}{l|c|c|c|c|c|c}
    \hline
         Question &       & H1     & H2 & H3 & H4 & H5 \\
          \hline
    \multirow{2}[0]{*}{What is the gross sale of Burger King} & Q & are   & \textbf{sales}, today & gross & is, what & burger, king  \\
          & A& based & sales, \textbf{14,billion}, 183 & diageo & contributed & burger, corp  \\
          \hline
    \multirow{2}[0]{*}{What is Florence Nightingale famous for} & Q & in, the & for   & \textbf{famous} & what  & florence, nightingale \\
          & A & of, in & was   & \textbf{nursing} & founder, modern, born & nightingale, italy   \\
          \hline
    \multirow{2}[0]{*}{Who is the founder of twitter?} & Q & the, of & -     & twitter, \textbf{founder} & - & who, is  \\
          & A & and, the & networking, launched & twitter, \textbf{jack dorsey}& match, social & -       \\
          \hline
    
    \end{tabular}%
    \caption{Analysis of QA pairs with respect to hierarchical level (H1-H5) based on vector norms. Self-organizing hierarchical structure facilitates better word level matching. Most informative word matches are marked in bold. Some words might be omitted from the answer due to lack of space. First two examples are from TrecQA and the third is from WikiQA. }
  \label{tab:big_table}%
\end{table*}
\begin{table}[htbp]

  \centering
  \small
    \begin{tabular}{ll}
    \hline
    $\norm{w}$  & Words (w) \\
    \hline
    0-1   & to, and, an, on, in, of, its, the, had, or, go \\
    1-2      & be, a, was, up, put, said, but  \\
      2-3    & judging, returning, volunteered, managing, meant, cited \\
      3-4    &  responsibility, engineering, trading, prosecuting\\
      4-5    & turkish, autonomous, cowboys, warren, seven, \textbf{what} \\
       5-6   & ebay, magdalena, spielberg, watson, nova  \\
       \hline
    \end{tabular}%


    \caption{Examples of words in each hierarchical level of the sphere based on vector norms. Smaller norms are closer to the core/origin.}
  \label{tab:word_norms}%
\end{table}%
\subsection{Analysis of QA Embeddings}

\begin{figure}[H]
\centering
\begin{subfigure}{0.24\textwidth}
  \centering
  \includegraphics[width=1.0\linewidth]{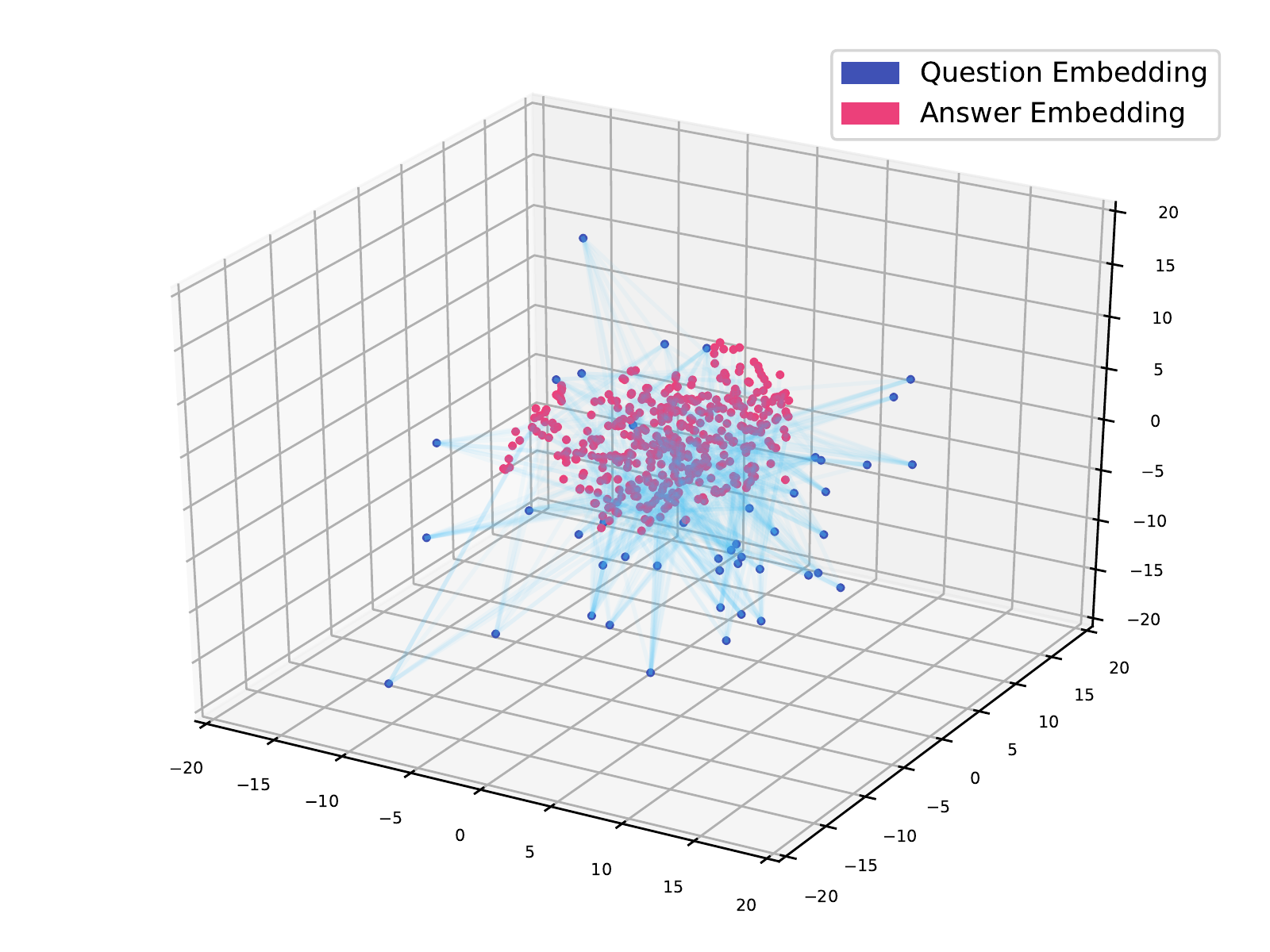}
  \caption{\textsc{HyperQA}}
  \label{fig:sub1}
\end{subfigure}%
\begin{subfigure}{0.24\textwidth}
  \centering
  \includegraphics[width=1.0\linewidth]{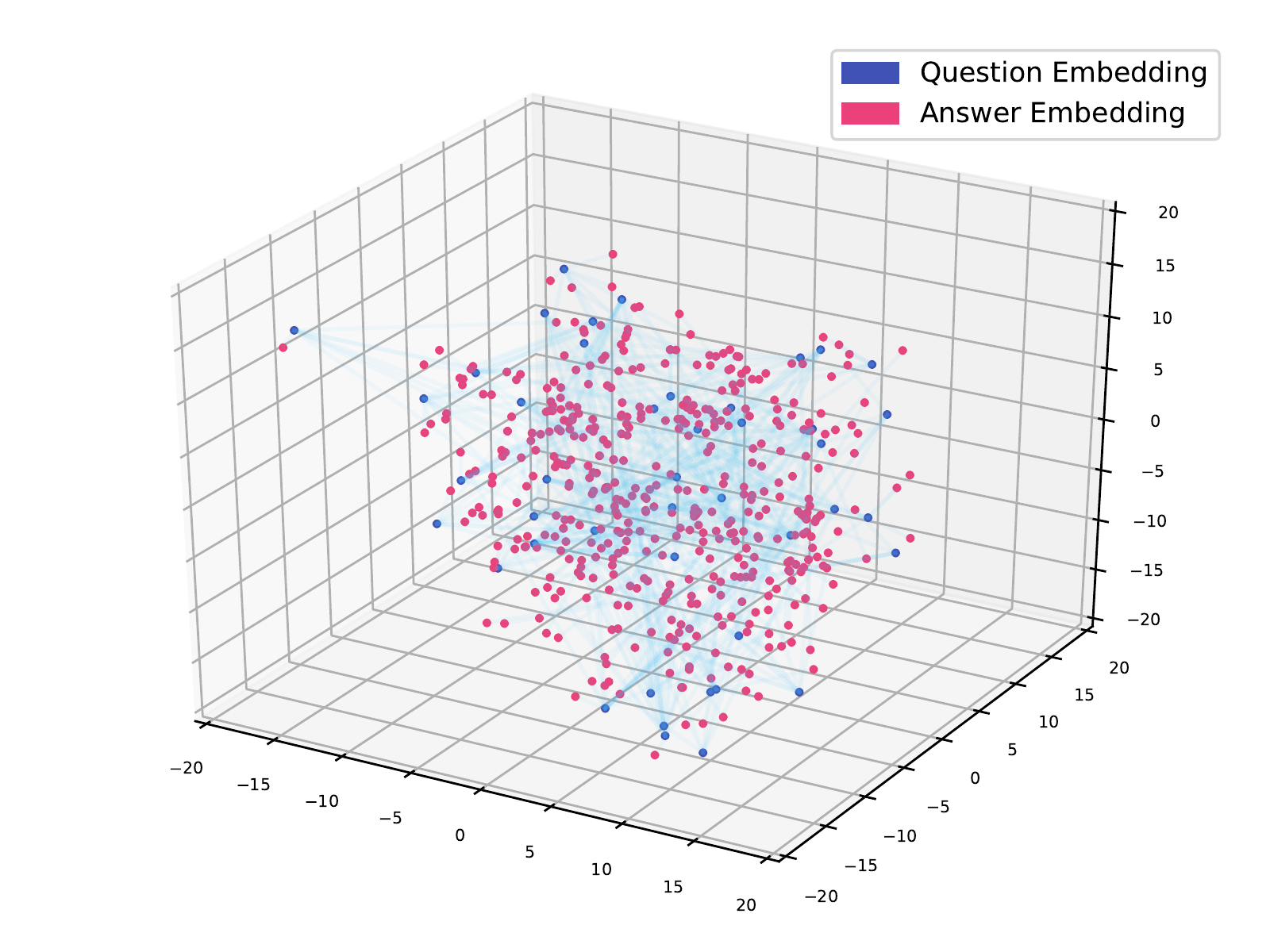}
  \caption{CosineQA}
  \label{fig:sub2}
\end{subfigure}
\caption{Visualization of QA embeddings on the test set of TrecQA. Projected to 3 dimensions using t-SNE. Blue lines depict mapping of question embeddings to their correct answers. \textsc{HyperQA} learns `sphere' shaped structure where questions embeddings surround the answer embeddings.}
\label{fig:viz_qa}
\end{figure}

\begin{figure}[H]
\centering

\begin{subfigure}{0.21\textwidth}
  \centering
  \includegraphics[width=1.0\linewidth]{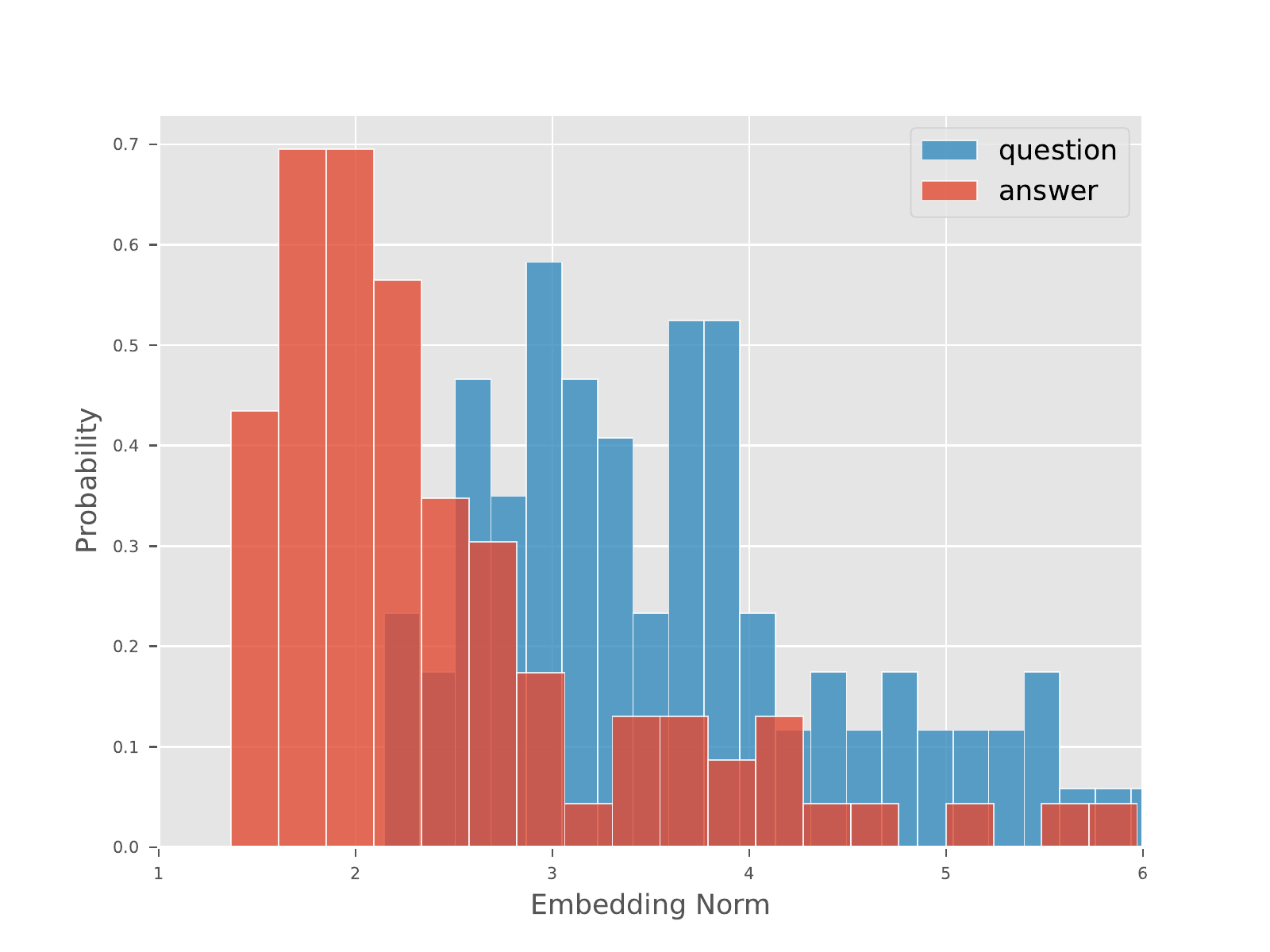}
  \caption{QA embeddings}
  \label{fig:sub3}
\end{subfigure}%
\begin{subfigure}{0.21\textwidth}
  \centering
  \includegraphics[width=1.0\linewidth]{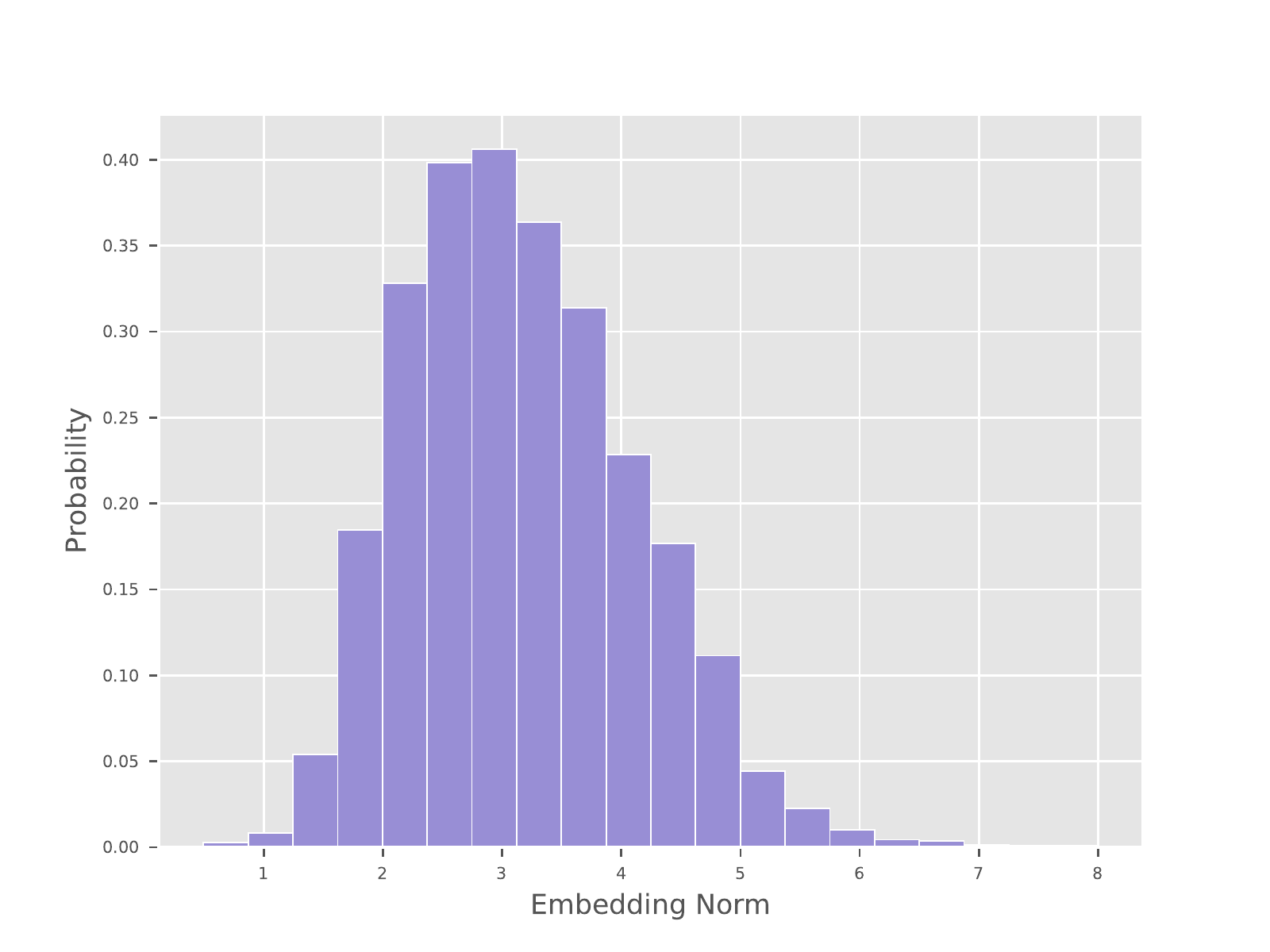}
  \caption{Word embeddings}
  \label{fig:sub4}
\end{subfigure}
\caption{Histogram plots of embedding norms.}
\label{fig:histograms}
\end{figure}

Figure \ref{fig:viz_qa}(a) shows a visualization of QA embeddings on the test set TrecQA projected in 3-dimensional space using t-SNE \cite{DBLP:journals/jmlr/Maaten09}. QA embeddings are extracted from the network as discussed in Section \ref{qa_sec}. We observe that question embeddings form a `sphere' over answer embeddings. Contrastingly, this is not exhibited when the cosine similarity is used as shown in Figure \ref{fig:viz_qa}(b). It is important to note that these are embeddings from the \textbf{test} set which have not been trained and therefore the model is not explicitly told whether a particular textual input is a question or answer. This demonstrates the innate ability of \textsc{HyperQA} to self-organize and learn latent hierarchies which directly answers \textbf{RQ1}. Additionally, Figure \ref{fig:histograms}(a) shows a histogram of the vector norms of question and answer embeddings. We can clearly see that questions in general have a higher vector norm\footnote{We extract QA embeddings right before the constraining / normalization layer.} and are at a different hierarchical level from answers. In order to further understand what the model is doing, we delve deeper into the visualization at word-level.  

\subsection{Analysis of Word Embeddings}
Table \ref{tab:word_norms} shows some examples of words at each hierarchical level of the sphere on TrecQA. Recall that the vector norms\footnote{Note that word embeddings are not constrained to $\norm{x}<1$.} allow us to infer the distance of the word embedding from the origin which depicts its hierarchical level in our context. Interestingly, we found that \textsc{HyperQA} exhibits self-organizing ability even at word-level. Specifically, we notice that the words closer to the origin are common words such as `to', `and' which do not have much semantic values for QA problems. At the middle of the hierarchy ($\norm{w} \approx 3$), we notice that there are more verbs. Finally, as we move towards the surface of the `sphere', the words become rarer and reflect more domain-specific words such as `ebay' and `spielberg'. Moreover, we also found many names and proper nouns occurring at this hierarchical level. 

\begin{figure}[H]
\begin{center}
\includegraphics[width=0.38\textwidth]{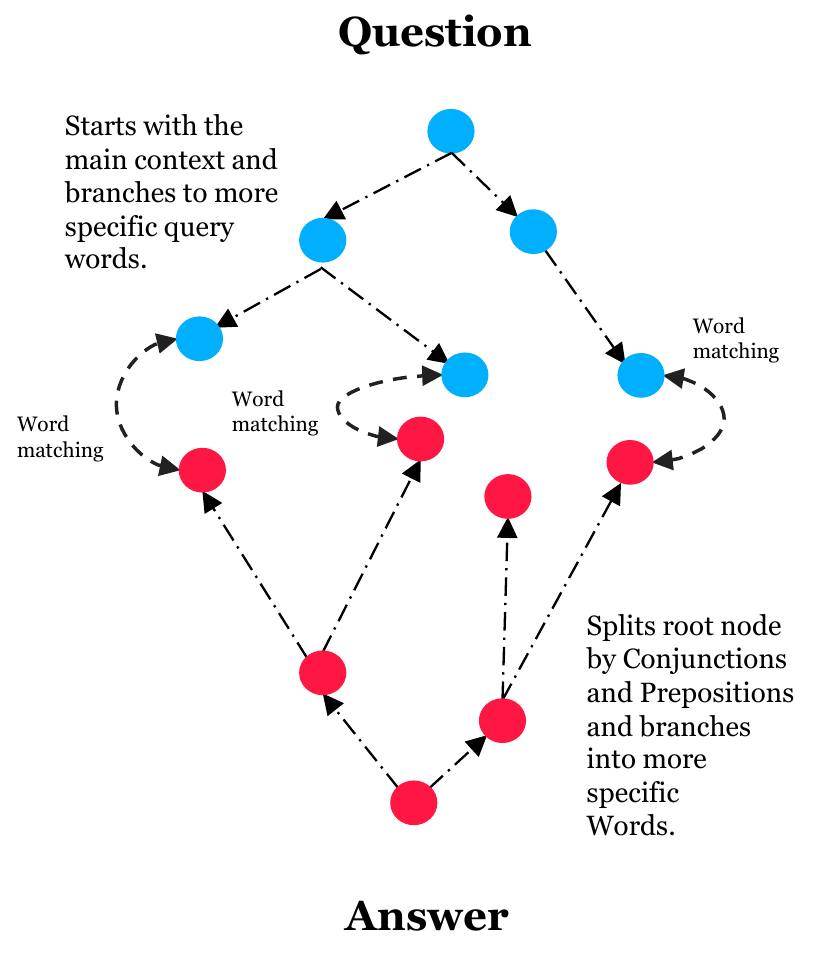}
\vspace{-1em}
\caption{Dual hierarchical matching of question and answer at word-level.}
\label{words_ex}
\end{center}
\end{figure}

 Additionally, we also observe that words such as 'where' or 'what' have relatively high vector norms and located quite high up in the hierarchy. This is in concert with Figure \ref{fig:viz_qa} which shows the question embeddings form a sphere around the answer embeddings. At last, we parsed QA pairs word-by-word according to hierarchical level (based on their vector norm). Table \ref{tab:big_table} reports the outcome of this experiment where $H1-H5$ are hierarchical levels based on vector norms. First, we find that questions often start with the overall context and drill down into more specific query words. Take the first sample in Table \ref{tab:big_table} for example, it begins at a top level with `burger king' and then drills down progressively to 'what is gross sales?'. Similarly in the second example, it begins with `florence nightingale' and drills down to `famous' at H3 in which a match is being found with `nursing' in the same hierarchical level. Overall, based on our qualitative analysis, we observe that, \textsc{HyperQA} builds two hierarchical structures at the \textbf{word-level} (in vector space) towards the middle which strongly facilitates word-level matching. Pertaining to answers, it seems like the model builds a hierarchy by splitting on conjunctive words (`and'), i.e., the root node of the tree starts by conjunctive words at splits sentences into semantic phrases. Overall, Figure \ref{words_ex} depicts our key intuitions regarding the inner workings of \textsc{HyperQA} which explains both \textbf{RQ2} and \textbf{RQ3}. This is also supported by Figure \ref{fig:histograms}(b) which shows the majority of the word norms are clustered with $\norm{w} \approx 3$. This would be reasonable considering that the leaf nodes of both question and answer hierarchies would reside in the middle. 

\vspace{-1em}

\section{Conclusion}
We proposed a new neural ranking model for question answering. Our proposed \textsc{HyperQA} achieves very competitive performance on four well-studied benchmark datasets. Our model is light-weight, fast and efficient, outperforming many state-of-the-art models with complex word interaction layers, attentive mechanisms or rich neural encoders. Our model only has 40K-90K parameters as opposed to millions of parameters which plague many competitor models. Moreover, we derive qualitative insights pertaining to our model which enable us to further understand its inner workings. Finally, we observe that the superior generalization of our model (despite small parameters) can be attributed to self-organizing properties of not only question and answer embeddings but also word embeddings.

\bibliographystyle{ACM-Reference-Format}
\bibliography{references} 

\end{document}